\documentstyle[prl,aps]{revtex}
\begin{document}
\draft
\title{Large Neutrino Mixing with Universal Strength of Yukawa Couplings}
\author{G. C. Branco\cite{gustavo}, M. N.\ Rebelo\cite{gui}}
\address{Centro de F\'\i sica das\\
Interac\c c\~oes Fundamentais (CFIF),\\
Instituto Superior T\'ecnico,\\
Av. Rovisco Pais, P-1096 Lisboa-Codex, Portugal}
\author{J. I. Silva-Marcos\cite{juca}}
\address{NIKHEF, Kruislaan 409, 1098 SJ Amsterdam, The Netherlands}
\maketitle

\begin{abstract}
We analyse, within the framework of universal strength for Yukawa couplings
(USY), various structures for the Dirac and Majorana neutrino mass matrices
giving rise, through the see-saw mechanism, to a degenerate mass spectrum. A
specific USY ansatz is presented for the charged lepton and neutrino
effective mass matrix, leading to quasi-degenerate neutrinos and a leptonic
mixing matrix which provides a large angle solution for both the atmospheric
and solar neutrino problems.
\end{abstract}

\pacs{14.60 Pq, 12.15Ff}


\subsection*{I.Introduction}

The measurement of solar and atmospheric neutrino fluxes provides
experimental evidence pointing towards neutrino oscillations, thus implying
non-zero neutrino masses and leptonic mixing. These exciting results have
motivated various attempts at understanding the structure of neutrino masses
and mixing \cite{ref1}. Assuming three neutrinos, the required neutrino mass
differences are such that in order for neutrinos to be of cosmological
relevance, their masses have to be approximately degenerate. For Majorana
neutrinos the case of quasi-degeneracy is specially interesting, since
mixing and CP violation can occur even in the limit of exact mass degeneracy 
\cite{ref30}.

In this paper, we propose a simple ansatz within the framework of universal
strength for Yukawa couplings (USY) \cite{ref2} which leads in a natural way
to a set of highly degenerate neutrinos, while providing a large mixing
solution for both the solar and atmospheric neutrino data. Within USY, all
Yukawa couplings have equal moduli, but different complex phases, thus
leading to complex unimodular mass matrices. We extend this idea to the
leptonic sector, choosing ans\"atze where the charged lepton and neutrino
mass matrices have this special form. In the quark sector the USY hypothesis
already proved to be quite successful, leading to ans\"atze for the Yukawa
couplings, where the parameters of the Cabibbo-Kobayashi-Maskawa matrix are
predicted in terms of quark mass ratios, without any free parameters \cite
{ref3}.

The most recent results of the SuperKamiokande (SK) collaboration \cite{ref4}
\cite{ref4a} strengthen the possibility of nearly maximal mixing angle for
atmospheric neutrino oscillations with the experimental parameters within
the range \cite{ref5} $\Delta m_{atm}^2=(1.5-8)\times 10^{-3}\ eV^2$, $\sin
^2(2\theta_{atm})>0.8$. In the absence of sterile neutrinos the dominant
mode is $\nu _\mu \longleftrightarrow \nu _\tau $ oscillations while the
sub-leading mode $\nu _\mu \longleftrightarrow \nu _e$ is severely
restricted by the SK and CHOOZ \cite{ref6} data which require $V_{13}\leq 0.2
$ for the range given above.

The interpretation of the present solar neutrino data leads to oscillations
of the electron neutrino into some other neutrino species, with three
different ranges of parameters still allowed. In the framework of the MSW
mechanism \cite{ref7} there are two sets of solutions, the adiabatic branch
(AMSW) requiring a large mixing, ($\Delta m_{sol}^2=(2-20)\times 10^{-5}\
eV^2$, $\sin ^2(2\theta _{sol})=0.65-0.95$) \cite{ref4a} \cite{ref8}, and
the non-adiabatic branch (NAMSW) requiring small mixing ($\Delta m_{sol}^2=
5.4\times 10^{-6}\ eV^2$, $\sin ^2(2\theta _{sol})\sim 6.0\times 10^{-3}$,
for the best fit) \cite{ref8}. In the framework of vacuum oscillations again
large mixing is required ($\Delta m_{sol}^2=8.0 \times 10^{-11}\ eV^2$, $%
\sin ^2(2\theta _{sol})= 0.75$, for the best fit) \cite{ref8}.

The result of the LSND collaboration based on a reactor experiment \cite
{ref11} has not yet been confirmed by other experiments and in particular
the KARMEN\ data \cite{ref12} already excludes a sizeable part of the
allowed parameter space. In this paper, we only take into consideration the
solar and atmospheric neutrino data and consider three neutrino families
without additional sterile neutrinos.

The paper is organized as follows. In the next section, we analyse various
possibilities for having a degenerate or quasi-degenerate neutrino mass
spectrum in USY, within the framework of the see-saw mechanism. In section
III, we present a specific USY ansatz for charged lepton and neutrino mass
matrices. In section IV, we show through numerical examples, how the ansatz
can accommodate all present data on atmospheric and solar neutrinos. Finally
our conclusions are presented in section V.

\subsection*{II. The See-saw mechanism and USY}

The see-saw mechanism provides one of the most attractive scenarios for
having naturally small masses for the left-handed neutrinos. Although the
mechanism has been introduced within the framework of models with an
extended gauge group such as $SO(10)$ \cite{ref11a} or $SU(2)_L\times
SU(2)_R\times U(1)$ \cite{ref11b}, it is clear that one may have the see-saw
mechanism within the standard $SU(3)_C\times SU(2)_L\times U(1)$ theory,
through the introduction of three right-handed neutrinos, with no other
modification. The full $(6\times 6)$ neutrino mass matrix can be written as: 
\begin{equation}
\label{eq0}{\cal M}\ =\ \left[ 
\begin{array}{ccc}
0 & m_D &  \\ 
m^T_D & M_R &  
\end{array}
\right]\ 
\end{equation}
where $m_D$ denotes the neutrino Dirac mass matrix, while $M_R$ stands for
the right-handed Majorana mass matrix. The Dirac mass matrix is proportional
to a vacuum expectation value $v$ of the Higgs doublet responsible for the $%
SU(2)\times U(1)$ breaking, while the right-handed Majorana mass term, being
invariant under $SU(2)\times U(1)$, has a scale $V_o$ which can be much
larger than $v$. The masses and mixing of the left-handed neutrinos are
determined by an effective mass matrix given by: 
\begin{equation}
\label{eq0a}m_{eff}\ =\ -m_D\ M^{-1}_R\ \ m^T_D 
\end{equation}

In this section, we analyse the various structures for $m_D$ and $M_R$,
which can lead to $m_{eff}$ corresponding to quasi-degenerate neutrinos. We
are specially interested in structures based on the USY principle. We will
consider various examples, without attempting at being exhaustive. For
simplicity, we will consider the exact degeneracy limit. The
quasi-degenerate case can be viewed as a small perturbation around that
limit.

Within the USY framework, exact mass degeneracy for a $3\times 3$ matrix is
achieved for a mass matrix proportional to $Y$, where, 
\begin{equation}
\label{eq0c}Y\ =\ \frac{1}{\sqrt{3}}\ \left[ 
\begin{array}{ccc}
\omega & 1 & 1 \\ 
1 & \omega & 1 \\ 
1 & 1 & \omega 
\end{array}
\right] 
\end{equation}
with $\omega = e^{2\pi i/3}$. It can be readily verified that $Y$ can also
be written as 
\begin{equation}
\label{eq0d}Y\ =\ e^{\frac{5\pi i}{6}}\ F\cdot {\text{diag}}(1,1,\omega^*)
\cdot F^T 
\end{equation}
where $F$ is given by 
\begin{equation}
\label{eq0e}F=\ \left[ 
\begin{array}{ccc}
\frac 1{\sqrt{2}} & \frac{-1}{\sqrt{6}} & \frac 1{\sqrt{3}} \\ \frac{-1}{%
\sqrt{2}} & \frac{-1}{\sqrt{6}} & \frac 1{\sqrt{3}} \\ 0 & \frac 2{\sqrt{6}}
& \frac 1{\sqrt{3}} 
\end{array}
\right] \ 
\end{equation}
In the framework of the see-saw mechanism, there are various cases which can
lead to mass degeneracy.

\paragraph*{Case I}

Both the Dirac and the right-handed Majorana mass matrices are proportional
to $Y$ so that one obtains for the full neutrino mass matrix: 
\begin{equation}
\label{eq0f}{\cal M}\ =\ \left[ 
\begin{array}{ccc}
0 & \lambda Y &  \\ 
\lambda Y & \mu Y &  
\end{array}
\right]\ 
\end{equation}
where $\lambda$, $\mu$ are real constants with dimension of mass, satisfying
the relation $\lambda^2/\mu\approx v^2/V_o$. The effective $3\times 3$ mass
matrix is then given by 
\begin{equation}
\label{eq0g}m_{eff}\ =\ -\frac{\lambda^2}{\mu}\ Y\ Y^{-1}\ Y= \ -\frac{%
\lambda^2}{\mu}\ Y 
\end{equation}
One concludes that if $m_D$ and $M_R$ are proportional to $Y$, then $m_{eff}$
will also have a degenerate mass spectrum.

\paragraph*{Case II}

Both $M_R$ and $M_D$ have again degenerate eigenvalues, but we assume that $%
M_R$ is proportional to $Y$ in the weak-basis where $m_D$ is already
diagonal and therefore proportional to the unit matrix. The neutrino mass
matrix has then the form: 
\begin{equation}
\label{eq0h}{\cal M}\ =\ \left[ 
\begin{array}{ccc}
0 & \lambda {\openone} &  \\ 
\lambda {\openone} & \mu Y &  
\end{array}
\right] 
\end{equation}
which leads to 
\begin{equation}
\label{eq0i}m_{eff}\ =\ -\frac{\lambda^2}{\mu}\ Y^{-1}\ = \ -\frac{\lambda^2%
}{\mu}\ Y^* 
\end{equation}
It is clear from Eq.(\ref{eq0d}) that $m_{eff}$ has also a degenerate mass
spectrum.

\paragraph*{Case III}

Let us now consider a situation analogous to case II, but where the forms of 
$M_R$ and $m_D$ are interchanged, i.e. 
\begin{equation}
\label{eq0j}{\cal M}\ =\ \left[ 
\begin{array}{ccc}
0 & \lambda Y &  \\ 
\lambda Y & \mu {\openone} &  
\end{array}
\right]\ 
\end{equation}
which implies 
\begin{equation}
\label{eq0k}m_{eff}\ =\ -\frac{\lambda^2}{\mu}\ Y^{2}\ = \ -\frac{\lambda^2}{%
\mu}\ i\ Y^* 
\end{equation}
so that one obtains again $m_{eff}$ with a degenerate mass spectrum.

\paragraph*{Case IV}

So far, we have only considered cases where both $m_D$ and $M_R$ have
degenerate eigenvalues. We shall now assume that $m_D$ has an hierarchical
spectrum and show that one may obtain a $m_{eff}$ with degenerate spectrum,
using a $M_R$ which has an hierarchical spectrum also. For definiteness, let
us assume that $m_D$ is given by 
\begin{equation}
\label{eq0l}m_D\ =\ \lambda \left[ 
\begin{array}{ccc}
e^{i \epsilon_1} & 1 & 1 \\ 
1 & e^{i \epsilon_1} & 1 \\ 
1 & 1 & e^{i \epsilon_2} 
\end{array}
\right] 
\end{equation}
where $\epsilon_i$ are real parameters, satisfying the relations $%
|\epsilon_1|<<|\epsilon_2|<<1$. The matrix $m_D=\lambda \ m_{D_o}$ can be
written as a sum with 
\begin{equation}
\label{eq0m}m_{D_o}\ =\ \Delta\ + \ \epsilon_1 A\ +\ \epsilon_2 B 
\end{equation}
where 
\begin{equation}
\label{eq0n}\Delta\ =\ \left[ 
\begin{array}{ccc}
1 & 1 & 1 \\ 
1 & 1 & 1 \\ 
1 & 1 & 1 
\end{array}
\right] 
\end{equation}
and 
\begin{equation}
\label{eq0o}
\begin{array}{cc}
A\ =\ \frac{(e^{i \epsilon_1}-1)}{\epsilon_1}\ \ \text{diag}(1,1,0) ,\quad & 
B\ =\ \frac{(e^{i \epsilon_2}-1)}{\epsilon_2}\ \ \text{diag}(0,0,1) 
\end{array}
\end{equation}
Since $A$, $B$ are of order $1$, it is clear that $m_D$ has a hierarchical
spectrum. If we choose now 
\begin{equation}
\label{eq0p}M_R\ =\ {\mu}\ \ m_{D_o}\ Y^{*}\ \ m_{D_o} 
\end{equation}
one obtains 
\begin{equation}
\label{eq0q}m_{eff}\ =\ -\frac{\lambda^2}{\mu}\ m_{D_o} \ \left[ m_{D_o}\
Y^{*}\ \ m_{D_o}\right]^{-1}\ m_{D_o}\ = \ -\frac{\lambda^2}{\mu}\ Y 
\end{equation}
where we have used the fact that $(Y^*)^{-1}=Y$. It is clear that $m_{eff}$
has a degenerate spectrum. The interesting point is that $M_R$ has a
hierarchical spectrum, since from Eqs.(\ref{eq0m}, \ref{eq0p}) one obtains 
\begin{equation}
\label{eq0r}M_R\ =\ {\mu}\  \left[\Delta+ \epsilon_1 A+\epsilon_2
B\right]\cdot Y^{*}\cdot \left[\Delta+ \epsilon_1 A+\epsilon_2 B\right] = \
3e^{-\frac{\pi i}{6}}\ {\mu}\ \left[\Delta+ \epsilon_1 A^{\prime}+\epsilon_2
B^{\prime}\right] 
\end{equation}
where we have used the fact that for any matrix $Z$, one has $\Delta\ Z\
\Delta=(\sum_{ij} Z_{ij})\ \Delta$. It is clear from Eq.(\ref{eq0r}) that $%
M_R$ has indeed a hierarchical spectrum since $A^{\prime}$ and $B^{\prime}$
are at most of order one.

We have shown that starting from a hierarchical Dirac neutrino mass $m_D$,
it is always possible to find a Majorana mass matrix $M_R$ which leads to an
exactly degenerate mass matrix of the USY type. However, it should be
stressed that in order to achieve that, it is required a significant amount
of fine-tuning between the Dirac and Majorana sectors, unless there is a
symmetry principle constraining both sectors.

\subsection*{III. A specific ansatz within the USY framework}

In this section, we suggest the following specific ansatz for the charged
lepton mass matrix $M_\ell$ and the effective $3\times 3$ neutrino mass
matrix $M_\nu$: 
\begin{equation}
\label{eq2}M_\ell = 
\begin{array}{cc}
c_\ell \ \left[ 
\begin{array}{ccc}
e^{ia} & 1 & 1 \\ 
1 & e^{ia} & 1 \\ 
1 & 1 & e^{-i(a+b)} 
\end{array}
\right] \ ,\quad \  & M_\nu =c_\nu \ \left[ 
\begin{array}{ccc}
e^{i\alpha } & 1 & 1 \\ 
1 & e^{i\beta } & 1 \\ 
1 & 1 & e^{-i(\alpha +\beta )} 
\end{array}
\right] \  
\end{array}
\end{equation}
Both $M_\ell $ and $M_\nu $ are of the USY type, symmetric and with only
three real free parameters each, thus leading to full calculability of the
mixing angles in terms of the mass ratios. Note that $M_\nu $ is the
relevant mass matrix for the neutrinos; it can either be an effective
see-saw mass matrix, as discussed in the previous section or simply a
Majorana mass matrix for left-handed neutrinos in a model with no
right-handed neutrinos.

The leptonic charged weak current interactions can be written as: 
\begin{equation}
\label{eq3}{\cal L}_W=\frac{g_{_W}}2\ (\overline{e},\overline{\mu },%
\overline{\tau })_L\ \ \gamma _\mu \ \ V\ \ \ \left( 
\begin{array}{c}
\nu _1 \\ 
\nu _2 \\ 
\nu _3 
\end{array}
\right) _L\quad W^\mu \quad +\quad \text{h.c.} 
\end{equation}
where the leptonic mixing matrix $V$ is given by: 
\begin{equation}
\label{eq4}V\ =\ U_\ell ^{\dagger }\cdot U_\nu 
\end{equation}
and where 
\begin{equation}
\label{eq5}
\begin{array}{cc}
\ell _{L_i}^{\text{weak}}\ =\ (U_\ell )_{ij}\ \ell _{L_i}^{\text{phys}}\
,\quad & \nu _{L_\alpha }\ =\ (U_\nu )_{\alpha i}\ \nu _{L_i} 
\end{array}
\end{equation}
with $\ell _{L_i}^{\text{phys}}$ denoting the physical charged leptons and $%
\nu_{L_i}$ the physical light neutrinos. The charged leptons have
hierarchical masses, thus implying that the phases $a$ and $b$ in Eq.(\ref
{eq2}) have to be small. These phases can be expressed in terms of the
charged lepton masses and to leading order one obtains: 
\begin{equation}
\label{eq6}
\begin{array}{cc}
|a|\ =\ 3\ \frac{m_e}{m_\tau }\ ,\quad \  & |b|\ =\ \frac 92\ \frac{m_\mu }{%
m_\tau } 
\end{array}
\end{equation}

On the other hand, we want the matrix $M_\nu $ in Eq.(\ref{eq2}) to lead to
highly degenerate neutrino masses. It can be easily checked that the matrix 
\begin{equation}
\label{eq7}M=c\ \left[ 
\begin{array}{ccc}
e^{i\alpha } & 1 & 1 \\ 
1 & e^{i\alpha } & 1 \\ 
1 & 1 & e^{-i2\alpha } 
\end{array}
\right] \ 
\end{equation}
has in general two degenerate eigenvalues and that in particular, for $%
\alpha =2\pi /3$ we recover the $Y$ matrix of Eq.(\ref{eq0c}), where all
three eigenvalues are exactly degenerate. This suggests that we expand $%
\alpha $ and $\beta $ in Eq.(\ref{eq2}) around the value $2\pi /3$,
introducing two small parameters $\delta $ and $\varepsilon $ defined by: 
\begin{equation}
\label{eq8}
\begin{array}{cc}
\alpha \ =\frac{2\pi }3-\delta -\varepsilon \ ,\quad \  & \beta \ =\frac{%
2\pi }3-\delta 
\end{array}
\end{equation}
In the limit $\varepsilon =0$, one still has a two-fold degeneracy of
eigenvalues, as in Eq.(\ref{eq7}). The eigenvalues $\lambda _i$ of the
dimensionless hermitian matrix $H_\nu \equiv (M_\nu \ M_\nu ^{\dagger
})/(3c_\nu ^2)$ are given in terms of $\alpha $ and $\beta $ by the
expression 
\begin{equation}
\label{eq9}\lambda _i\ =1+\ 2\ x\ \cos \phi _i 
\end{equation}
with 
\begin{equation}
\label{eq10}
\begin{array}{ccc}
\phi _1\ =\theta +\frac{\alpha -\beta }3-\frac{2\pi }3\ \ ,\quad \  & \phi
_2\ =\theta +\frac{\alpha -\beta }3+\frac{2\pi }3\ \ ,\quad & \phi _3\
=\theta +\frac{\alpha -\beta }3 
\end{array}
\end{equation}
and 
\begin{equation}
\label{eq11}
\begin{array}{cc}
\tan (\theta )\ =\frac{\sin \beta -\sin \alpha }{\cos \beta +\cos \alpha +1}%
\ ,\quad \  & x\ =\frac 13\sqrt{3+2\cos (\beta +\alpha )+2\cos \beta +2\cos
\alpha } 
\end{array}
\end{equation}
The parameters $\delta $ and $\varepsilon $ can be expressed in terms of
neutrino masses, and in leading order one has: 
\begin{equation}
\label{eq12}
\begin{array}{cc}
|\delta| \ =\frac 1{\sqrt{3}}\ \ \frac{\Delta m_{32}^2}{m_3^2}\ ,\quad \  & 
|\varepsilon| \ =\sqrt{3}\ \ \frac{\Delta m_{21}^2}{m_3^2} 
\end{array}
\end{equation}
where $\Delta m_{ij}^2=|m_i^2-m_j^2|$. The matrix $H_\ell \equiv (M_\ell \
M_\ell ^{\dagger })/(3c_\ell ^2)$ is approximately diagonalized by $U_\ell =F
$ defined in Eq.(\ref{eq0e}), with additional small corrections expressible
in terms of charged lepton mass ratios. The diagonalization of $M_\nu $
requires special care since to leading order $M_\nu $ is an exactly
degenerate mass matrix. In Ref. \cite{ref30} we have studied the general
form of Majorana neutrino mass matrices leading to exact degeneracy and we
have pointed out that if a given unitary matrix $U_\circ$ diagonalizes the
degenerate mass matrix, so does the matrix $U_\nu=U_\circ \ O$, with $O$ an
arbitrary orthogonal matrix. The diagonalizing matrix $U_\nu=U_\circ \ O$ is
only fixed when the mass degeneracy is lifted. For our specific case with $%
M_\nu$ given by Eq.(\ref{eq2}), we obtain in next to leading order : 
\begin{equation}
\label{eq14}U_\nu =\frac {e^{-\frac{\pi i}{4}}}{\sqrt{3}}\ \left[ 
\begin{array}{ccc}
\omega & 1 & 1 \\ 
1 & \omega & 1 \\ 
1 & 1 & \omega 
\end{array}
\right] \ \cdot K 
\end{equation}
where $K=$ diag $(-1,1 , 1)$, so that $U_{\nu}^{T} M_{\nu} U_{\nu}$ is
diagonal real and positive for a positive $c_{\nu}$. As a result the moduli
of the mixing matrix are, to a very good approximation, given by: 
\begin{equation}
\label{eq15}|V| \simeq \ \left[ 
\begin{array}{ccc}
\frac 1{\sqrt{2}} & \frac 1{\sqrt{2}} & 0 \\ 
\frac{1}{\sqrt{6}} & \frac{1}{\sqrt{6}} & \frac 2{\sqrt{6}} \\ \frac 1{\sqrt{%
3}} & \frac 1{\sqrt{3}} & \frac 1{\sqrt{3}} 
\end{array}
\right] \ 
\end{equation}

\subsection*{IV. Confronting the data}

There is a stringent bound on the parameter $c_\nu$ of the neutrino mass
matrix in Eq.(\ref{eq2}) from neutrinoless double beta decay, which can be
expressed by $|<m>|\equiv |\sum_{i} U^2_{ei}m_{\nu_i}| = |m_{ee}| \ <\ 0.2\
\ eV$ \cite{ref15},with $m_{ee}$ denoting the entry $(11)$ of $M_{\nu}$ in
the weak basis where $M_{l}$ is diagonal. Taking into account Eq.(\ref{eq2}%
), this immediately leads to 
\begin{equation}
\label{eq16}m\ \ <\ \ \sqrt{3}\ \ c_\nu\ \approx\ 0.2\ \ eV 
\end{equation}
so that in the case of almost degenerate neutrinos coming from the ansatz of
Eq.(\ref{eq2}), we cannot have light neutrinos with masses higher than about 
$0.2\ eV$, where $m$ is the approximate neutrino mass.

In order to compare our ansatz with the experimental results from
atmospheric and solar neutrino experiments, we must bear in mind that in the
context of three left-handed neutrinos the probability for a neutrino $%
\nu_\alpha$ to oscillate into other neutrinos is given by 
\begin{equation}
\label{eq17}1-\text{P}(\nu_\alpha\rightarrow\nu_\alpha)\ =\ 4\ \sum_{i<j}\
|V_{\alpha i}|^2\ |V_{\alpha j}|^2\ \sin^2\left[ \frac {\Delta m_{ij}^2}{4}\ 
\frac L{E}\right] 
\end{equation}
where $E$ is the neutrino energy, and $L$ denotes the distance travelled
between the source and the detector. The translation of the experimental
bounds, which are given in terms of only two flavour mixing, into the three
flavour mixing is simple, since in this case we have $V_{13}$ close to zero
and also $\Delta m^2_{32}>> \Delta m^2_{21}$, and we may safely identify: 
\begin{equation}
\label{eq18}\sin^2 2\theta_{\text{atm}}\ =\ 4\ \left(|V_{21}|^2\ |V_{23}|^2\
+\ |V_{22}|^2\ |V_{23}|^2\ \right) 
\end{equation}
\begin{equation}
\label{eq19}\sin^2 2\theta_{\text{sol}}\ =\ 4\ |V_{11}|^2\ |V_{12}|^2 
\end{equation}

The following examples illustrate how our ansatz fits the experimental
bounds for large solar and atmospheric mixing. The first example is in the
context of vacuum oscillations and the second for large mixing AMSW.

\paragraph*{1st Example}

We choose as input the masses for the charged leptons 
\begin{equation}
\label{eq20}
\begin{array}{ccc}
m_e\ =\ 0.511\ MeV\ ,\quad & m_\mu\ =\ 105.7\ MeV\ ,\quad & m_\tau\ =\ 1777\
MeV 
\end{array}
\end{equation}
which correspond to phases $|a|= 8.61\times 10^{-4}$ and $|b|=0.267 $ of Eq.(%
\ref{eq2}). For the neutrino sector we choose 
\begin{equation}
\label{eq21}
\begin{array}{ccc}
m_{\nu_3}\ =\ 0.2\ eV\ ,\quad & \Delta m^2_{32}\ =\ 5.0\times 10^{-3}\ eV^2\
,\quad & \Delta m^2_{21}\ =\ 1.0\times 10^{-10}\ eV^2 
\end{array}
\end{equation}
thus fixing the parameters $|\delta|=0.0772 $ and $|\varepsilon|=4.98\times
10^{-9}$ of Eq.(\ref{eq8}).

Performing an exact numerical diagonalization of the mass matrices we obtain
for the leptonic mixing matrix 
\begin{equation}
\label{eq22}|V|=\ \left[ 
\begin{array}{ccc}
0.707 & 0.707 & 6.78\times 10^{-10} \\ 
0.406 & 0.406 & 0.819 \\ 
0.579 & 0.579 & 0.574 
\end{array}
\right] \ 
\end{equation}
which from Eq.(\ref{eq18}) and Eq.(\ref{eq19}) translates into 
\begin{equation}
\label{eq23}
\begin{array}{cc}
\sin^2(2\theta_{\text{atm}})\ =\ 0.884,\quad & \sin^2(2\theta_{\text{sol}})\
=\ 1.0 
\end{array}
\end{equation}

\paragraph*{2nd Example}

In this second numerical application, we choose 
\begin{equation}
\label{eq24}
\begin{array}{ccc}
m_{\nu_3}\ =\ 0.2\ eV\ ,\quad & \Delta m^2_{32}\ =\ 5.0\times 10^{-3}\ eV^2\
,\quad & \Delta m^2_{21}\ =\ 5.0\times 10^{-5}\ eV^2 
\end{array}
\end{equation}
in agreement with the large mixing AMSW solution for the solar problem. This
case corresponds to $|\delta|=0.0764 $ and $|\varepsilon|=2.48 \times 10^{-3}
$. The resulting leptonic mixing matrix coincides, to an excellent
approximation, with that of Eq.(\ref{eq22}), with the exception of $|V_{13}|$
which is given by $|V_{13}|=3.38 \times 10^{-4}$. Of course this is to be
expected from the discussion of section III where we have pointed out that
this ansatz implies in leading order a leptonic mixing matrix given by Eq.(%
\ref{eq15}). The resulting values for $\sin^2(2\theta_{\text{atm}})$ and $%
\sin^2(2\theta_{\text{sol}})$ do not deviate from those of Eqs.(\ref{eq23}).

In these examples, we fixed the parameters of our ansatz in such a way that
we reproduce the charged leptonic masses and obtain almost degenerate
neutrino masses obeying the current experimental bounds on neutrino mass
splitting. The ansatz then leads to large values for $\sin ^2(2\theta _{%
\text{atm}})$ and $\sin ^2(2\theta _{\text{sol}})$. Comparing our results
with the experimental constraints, we conclude that our ansatz is in better
agreement with the vacuum oscillation solution for solar neutrinos than with
AMSW solution since the value $\sin ^2(2\theta _{\text{sol}})=1.0$ is
disfavoured in the framework of AMSW \cite{ref20}. However, it should be
noted that it is possible, within USY, to obtain a $\sin ^2(2\theta _{\text{%
sol}})$ compatible with the AMSW solution, with a slight modification of our
ansatz, by replacing $M_\nu $ in Eq.(\ref{eq2}) by $M_\nu ^{\prime }=C\cdot
M_\nu \cdot C$, where $C=$diag$(1,e^{i\alpha },1)$. In this case, we obtain,
for $\alpha =-0.2$, $\sin ^2(2\theta _{\text{sol}})=0.974$, $\sin ^2(2\theta
_{\text{atm}})=0.934$, with the same mass splittings as before. This result
is inside the $90\%$ C.L. experimental value for the $\sin ^2(2\theta _{%
\text{sol}})$ in AMSW. 

Concerning the stability of our ansatz under the renormalization group
equations (RGE), we find that the VO solution is unstable, since the
required $\Delta m_{21}^2$ mass splitting, of the order of $10^{-10}\ eV^2$,
is much smaller than the mass splitting $\Delta m_{RGE}^2$ generated by the
running of the RGE. Indeed, in the framework of the standard model, one
finds \cite{ref21}: $\Delta m_{RGE}^2\approx m_\nu ^2\ \epsilon $, with
$\epsilon $
given by,

\begin{equation}
\label{eq24a}\epsilon \ =\ \frac{Y_\tau ^2}{32\pi ^2}\log \left( \frac 
\Lambda {M_Z}\right) 
\end{equation}
where $Y_\tau $ is the $\tau $ Yukawa coupling (at $M_Z$). It is clear, that
even for $\Lambda =o(10\ TeV)$, $\epsilon \geq o(10^{-6})$. Therefore,
with our $m_\nu =0.2\ eV$, the mass splitting coming from the RGE is always
larger that $10^{-8}\ eV^2$, and this far exceeds the required VO mass
splitting of $10^{-10}\ eV^2$. This result is in agreement with previous
analysis found in the literature \cite{ref21}, and it implies that in order
to have a stable VO solution, one would need a mechanism, like e.g. an exact
symmetry, which would protect $\Delta m_{21}^2$ from becoming too large. 

On the other hand, the MSW solution is quite stable. Even if we take $%
\Lambda =o(10^{19})\ GeV$, we get $\epsilon \leq 10^{-5}$, leading to a
RGE mass splitting of $\Delta m_{RGE}^2\approx 10^{-7}\ eV^2$, which is much
smaller than the required AMSW splitting of $\Delta m_{21}^2=5\times
10^{-5}\ eV^2$. The mixing angles are not significantly altered by the
running of the RGE.

\subsection*{V. Conclusions}

Within the framework of the USY hypothesis, we have analysed various
structures for the Dirac and Majorana neutrino mass matrices which can lead,
through the see-saw mechanism, to an effective neutrino mass matrix for the
left-handed neutrinos, with a degenerate mass spectrum. The physically
relevant case of quasi-degenerate neutrinos can be viewed as a small
perturbation of this limit. In one of the cases considered, the neutrino
Dirac mass matrix has a hierarchical spectrum, but the resulting effective
neutrino mass matrix has a degenerate spectrum. This case has the attractive
feature of having all fermions, namely quarks, charged leptons and neutrinos
with hierarchical Dirac masses. We have then put forward an USY ansatz for
the charged lepton and neutrino effective mass matrix, which leads to three
quasi-degenerate neutrinos . The ansatz is highly predictive since the
leptonic mixing matrix is given in leading order by a fixed matrix
(independent of the lepton masses) with small corrections given in terms of
lepton mass ratios, with no arbitrary parameters. A large mixing solution is
obtained both for the solar and atmospheric neutrino data.

We have verified that the VO solution is unstable under the RGE, while the
AMSW is stable. However, it should be pointed out that the problem of
stability cannot be separated from the question of obtaining the USY
structure from a symmetry principle. At present, this is still an open
question, and therefore our ansatz should be viewed as an effective theory
at low energies, resulting hopefully from an appropriate structure for the
Dirac and right-handed Majorana neutrinos, imposed at a high energy scale.

One of the salient features of the ansatz of Eq.(\ref{eq2}) is the fact that
the mass matrices for both the charged leptons and the neutrinos have
analogous structures, with all matrix elements with equal modulus and the
non-vanishing phases appearing only along the diagonal. The drastic
difference between the resulting spectra for the charged leptons and the
neutrinos has to do with the fact that in the case of charged leptons the
phases along the diagonal are small, while in the case of neutrinos the
phases are close to $2\pi /3$ which corresponds to the exact degeneracy
limit. These simple structures for the mass matrices and Yukawa couplings do
suggest the existence of a symmetry principle leading to them.

\begin{acknowledgements}
We thank E. Akhmedov for useful discussions.
M. N. R. and J. I. S. are thankful for the hospitality of CERN Theory
Division, where part of this work was done.
The work of J. I. S. was partially supported by Funda\c c\~ao
para a Ci\^encia e Tecnologia of the Portuguese Ministry of Science and
Technology.
\end{acknowledgements}


\begin{references}
\bibitem[*]{gustavo}  E-mail address: d2003@beta.ist.utl.pt

\bibitem[\dagger]{gui}  E-mail address: rebelo@beta.ist.utl.pt

\bibitem[\ddagger]{juca}  On leave from CFIF. E-mail address: juca@nikhef.nl

\bibitem{ref1}  V. Barger, S. Pakvasa, T. J. Weiler and K. Whisnant, Phys.
Lett. B 437 (1998) 107; G. C. Branco, M. N. Rebelo and J. I. Silva-Marcos,
Phys. Lett. B 428 (1998) 136; G. Altarelli and F. Feruglio, Phys. Lett. B
439 (1998) 112; {\it ibid} Phys. Lett. B 451 (1999) 388; M. Fukugita, M.
Tanimoto and T. Yanagida, Phys. Rev. D 57 (1998) 4429; H. Fritzsch and Z. Z.
Xing, Phys. Lett. B 440 (1998) 299; R. N. Mohapatra and S. Nussinov, Phys.
Lett. B 441 (1998) 299; {\it ibid} Phys. Rev. D 60 (1999) 013002; E. Ma,
Phys. Lett. B 442 (1998) 238; C. Wetterich, Phys. Lett. B 451 (1999) 397; J.
I. Silva-Marcos, Phys. Rev. D 59 (1999) 091301; F. Vissani, hep-ph/9708483;
H. Georgi and S. L. Glashow, hep-ph/9808293; J. A. Casas, V. Di Clemente A.
Ibarra and M. Quiros, hep-ph/9904295; M. Jezabek and Y. Sumino,
hep-ph/9904382; J. A. Casas, J. R. Espinosa, A. Ibarra and I. Navarro,
hep-ph/9905381.

\bibitem{ref30}  G. C. Branco, M. N. Rebelo and J. I. Silva-Marcos, Phys.
Rev. Lett. 82 (1999) 683.

\bibitem{ref2}  G. C. Branco, M. N. Rebelo and J. I. Silva-Marcos, Phys.
Lett. B 237 (1990) 446.

\bibitem{ref3}  G. C. Branco and J. I. Silva-Marcos, Phys. Lett. B 359
(1995) 166; G. C. Branco, D. Emmanuel-Costa and J. I. Silva-Marcos, Phys.
Rev. D 56 (1997) 107.

\bibitem{ref4}  SuperKamiokande collaboration, Y. Fukuda et al. Phys. Lett.
B 433 (1998) 9; {\it ibid} Phys. Lett. B 436 (1998) 33; {\it ibid} Phys.
Rev. Lett. 81 (1998) 1562;

\bibitem{ref4a}  SuperKamiokande collaboration, Y. Suzuki et al., talk given
at 17th International Workshop on Weak Interactions and Neutrinos (WIN'99),
24-30 January 1999, Cape Town, South Africa.

\bibitem{ref5}  O. L. Peres and A. Yu. Smirnov, hep-ph/9902312.

\bibitem{ref6}  CHOOZ collaboration, M. Apollonio et al., Phys. Lett. B 420
(1998) 397.

\bibitem{ref7}  L. Wolfenstein, Phys. Rev. D 17 (1978) 2369; {\it ibid} D 20
(1979) 2634; S. P. Mikheyev and A. Yu. Smirnov, Sov. J. Nucl. Phys. 42
(1985) 913; {\it ibid} Nuovo Cim. 9C (1986) 17; V. Barger et al., Phys. Rev.
D 22 (1980) 2718; H. A. Bethe, Phys. Rev. Lett. 56 (1986) 1305; S. P. Rosen
and J. M. Gelb, Phys. Rev. D 34 (1986) 969; J. Bouchez et al., Z. Phys. C 32
(1986) 499.

\bibitem{ref8}  J. N. Bahcall, P. I. Krastev and A. Yu. Smirnov, Phys. Rev.
D 58 (1998) 096016.

\bibitem{ref11}  LSND collaboration, C. Athanassopoulos et al. Phys. Rev.
Lett. 75 (1995) 2650; {\it ibid} Phys. Rev. Lett. 77 (1996) 3082; {\it ibid}
Phys. Rev. C 54 (1996) 2685; {\it ibid} Phys. Rev. Lett. 81 (1998) 1774; 
{\it ibid} Phys. Rev. C 58 (1998) 2489.

\bibitem{ref12}  KARMEN collaboration, B. Armbruster et al. Phys. Rev. C 57
(1998) 3414; G. Drexlin, talk at Wein'98, Santa Fe, June 14-21, 1998; K.
Eitel et al., 18th Int. Conf. on Neutrino Physics and Astrophysics (NEUTRINO
98), Takayama, Japan, 4-9 June 1998.

\bibitem{ref11a}  M. Gell-Mann, P. Ramond and R. Slansky, in Supergravity,
ed. by P. van Nieuwenhuizen and D. Z. Freedman (North Holland, Amsterdam,
1979), p.315; T. Yanagida, in Proc. of the Workshop on the Unified Theory
and Baryon Number in the Universe, ed. by O. Sawada and A. Sugamoto (KEK
report 79-18, 1979), p.95, Tsukuba, Japan.

\bibitem{ref11b}  R. Mohapatra and G. Senjanovic, Phys. Rev. Lett. 44 (1980)
912.

\bibitem{ref15}  L. Baudis et al., hep-ex/9902014.

\bibitem{ref20}  G. L. Fogli, E. Lisi and D. Montanino, Astropart. Phys. 9
(1998) 119.

\bibitem{ref21}  K. Hagiwara and N. Okamura, Nucl. Phys. B 548 (1999) 60; J.
Ellis and S. Lola, Phys. Lett. B 458 (1999) 310; J.A. Casas, J.R. Espinosa,
A. Ibarra and I. Navarro, JHEP 9909 (1999) 015; {\it ibid} hep-ph/9905381; 
{\it ibid} Nucl. Phys. B 556 (1999) 3; R. Barbieri, G. G. Ross and
A.Strumia, hep-ph/9906470; N. Haba, and N. Okamura, hep-ph/9906481; E. Ma,
hep-ph/9907400.
\end{references}
\end{document}